\documentstyle[12pt]{article}
\title{{Reply to ``A Comment on Singularities in  Quantum Cosmology''}
\thanks{This work is supported in part by funds provided by the U. S.
Department of Energy(D.O.E.) under cooperative research agreement
 \# DF-FC02-94ER40818 .}}
\author{\sc{Nivaldo A. Lemos}\thanks{
Supported by Conselho Nacional de Desenvolvimento Cient\'{\i}fico e
Tecnol\'ogico(CNPq), Brazil.}\\                                                
\small{\it Center for Theoretical Physics}\\
\small{\it Laboratory for Nuclear Science
and Department of Physics}\\
\small{\it Massachusetts Institute of Technology}\\
\small{\it Cambridge, MA 02139 - USA}\\
\small{\it and}\\
\small{\it Departamento de F\'{\i}sica, Universidade Federal Fluminense}\\
\small{\it Av. Litor\^anea s/n, Boa Viagem,  24210-340 Niter\'oi, RJ, Brazil}\\
\\
\small{e-mail: lemos@mitlns.mit.edu}
\\
\\
\\
\date{} }

\begin{document}
\pagestyle{myheadings} 
\baselineskip 18pt
 
%\markboth{L.V.Belvedere  }{Supersymmetric Liouville Theory}
\maketitle
\begin{abstract}

We argue that the reasonings that underlie a recent comment by Gotay and
Demaret [1] are defective. We maintain that, contrary to what they
assert, our previous papers [2,3,4] are correct and indeed disprove their 
conjecture
that quantum cosmological singularities are predetermined on the classical level
by the choice of time.                                                         
\vfill
\noindent CTP \# 2535 \hfill May 1996
\end{abstract}
\newpage

\noindent{\bf I. INTRODUCTION}
\vspace{0.75cm}

A significant problem of quantum cosmology is whether classical
singularities persist in the quantum domain, and how
quantum singularities are connected to the issue of time. 
In 1983, after  classifying the time variable $t$ of
a classically singular model as either ``slow'', if the classical singularity
occurs at a finite value of $t$,  or ``fast'', if the classical singularity
occurs at $t=\pm\infty$, Gotay and Demaret [5] 
conjectured that (F) {\it self-adjoint quantum dynamics in a fast-time gauge
is always singular}, whereas  (S) {\it self-adjoint quantum dynamics in a 
slow-time gauge is always nonsingular}.

By means of counterexamples, in [2,3,4] we disproved both of these conjectures.
In a recent Comment,
Gotay and Demaret [1] claim that these papers are incorrect and that their 
conjectures
 remain valid
for the quantum cosmologies considered in [2,3,4]. In this Reply we argue that
their objections are defective, and maintain that our previous results are
correct and indeed prove that both of their conjectures are false.

\vspace {1.25cm}

\noindent{\bf II. FRW MODEL WITH SCALAR FIELD}

\vspace {.75cm}

The model we will be dealing with is
a  spatially-flat Friedmann-Robertson-Walker universe
filled with a massless scalar field. The general model  was  originally 
introduced by Blyth and Isham [6], and the special case 
with $k=0$ and $m=0$ was considered  in [4], to
which the reader is directed for further details. Einstein's  ``$G_{00}$ 
equation''  is

$$3\frac{{\dot{R}}^2}{R^2} = \frac{1}{4}{{\dot{\phi}}^2 } \,\,\, ,\eqno(1)$$
\\                                                                          
\noindent where $R$ is the scalar factor and $\phi (t)$ is the homogeneous
scalar field. In the gauge $t=\phi$ the above equation is equivalent to

$$\dot{R} =   \left\{ \begin{array}{cl}
              R/{\sqrt {12}} & \mbox{if ${\dot R}>0$} \\
              \\
              - R/{\sqrt {12}} & \mbox{if ${\dot R}<0$}
              \end{array} \right.   \,\,\, .    \eqno(2)$$             
\\
\\
\noindent The  field equations allow of expanding or contracting 
universes,
that is, $R(t) = R_0 \exp (\pm t/{\sqrt{12}})$, where $R_0$ is an arbitrary
positive constant. These are mutually exclusive solutions, depending on the
initial conditions. The model is singular at  $t=-\infty$ in the expanding case
or at  $t=+\infty$ in the contracting case. Thus $t=\phi$ is a fast time.

The canonical momentum conjugate to $R$ is

$$p_R = \frac{\partial L}{\partial {\dot R}} =  12\frac{R{\dot R}}{N}
\eqno(3)$$
\\
\noindent whereas the momentum conjugate to $\phi$ is

$$p_{\phi} = \frac{\partial L}{\partial {\dot \phi}} = 
- \frac{R^3}{N}{\dot \phi} \,\,\, , \eqno(4)$$
\\                                                                       
\noindent where $N$ is the lapse. The super-Hamiltonian constraint is

$$ \frac{p_R^2}{24R} - \frac{p_{\phi}^2}{2R^3}  = 0 \,\,\, . \eqno(6)$$
\\                                                                             

According to the Arnowitt-Deser-Misner  reduction prescription, given the
choice $t=\phi$
the Hamiltonian in the reduced phase space is $H = - p_{\phi}$. Now,
solving Eq.(6) for $p_{\phi}$  and picking up the
negative square-root gives rise to the reduced Hamiltonian

$$ H  = \frac{1}{\sqrt{12}}\, R\, \vert p_R \vert 
\,\,\, . \eqno(6)$$
\\                                                             

\noindent The positive solution for
$p_{\phi}$ was discarded because in the gauge $t=\phi$
it follows from Eq.(4) that $p_{\phi}<0$ since $R>0$ and $N>0$
{\it by definition}. One sees, therefore, that the Hamiltonian
(6) is {\it naturally} positive. Hamilton's equation of motion  for $R$ 
in the reduced phase space  is

$${\dot R} = \frac{\partial H}{\partial p_R} =
              \left\{ \begin{array}{cl}
              R/{\sqrt {12}} & \mbox{if $p_R > 0$} \\
              \\
              - R/{\sqrt {12}} & \mbox{if $p_R < 0$}
              \end{array} \right. \,\,\, .      \eqno(7)$$             
\\
\\
\noindent Again because $R>0$ and $N>0$  by definition, it is a consequence of 
Eq.(3) that $p_R$ and $\dot{R}$ have the same sign, so that Eqs.(2) and (7) 
are identical.
This completes the verification that the equations of motion generated by the
reduced Hamiltonian (6) are completely equivalent to Einstein's equations for 
the gravitational field coupled to the scalar field. Such a consistency check 
is indispensable if  minisuperspace quantization
 is to have any meaning at all.      

Gotay and Demaret [1], however, take for Hamiltonian in the reduced phase space
                                                                               
$$ H_{GD}  = \frac{1}{\sqrt{12}}\, R\,  p_R \,\,\, , \eqno(8)$$
\\                                                  
\noindent and with the help of a standard symmetrization procedure write down
the corresponding Hamiltonian operator

$${\hat H}_{GD} = \frac{-i}{\sqrt{12}} \Bigl ( R\frac{d}{dR} + \frac{1}{2}
\Bigr ) \,\,\, . \eqno(9)$$
\\           
\noindent Notice that the Hamiltonian (8) generates only {\it half} of 
Einstein's
equations, that is, it excludes contracting universes. Therefore, it {\it is
not} the correct reduced Hamiltonian, and the results  based
on its quantum counterpart (9) that Gotay and Demaret  obtain in Section 2 
of their Comment are irrelevant because they
refer to something different from the model quantized in [4].

It is not difficult to trace their error.  The reduced phase space 
$ (R,p_R)$ is the union 
of the two disjoint sets  
$ (0,\infty ) \times (0,\infty )$ and
$ (0,\infty ) \times (-\infty ,0)$. If $p_R>0$ the universe expands from a 
singularity at $t=-\infty$, whereas  if 
$p_R<0$ the universe contracts to a singularity at $t=+\infty$, and {\it at the
classical level} these are mutually exclusive situations. Gotay and
Demaret  consider only the first case, and
supposing $p_R>0$ they are led to their Hamiltonian (8). The crux of the
matter lies in their innocent-looking phrase ``we will consider only the first
case.'' In so doing they in fact discuss a different model, that is, the
original model subject to the {\it additional constraint} $p_R>0$. The
original classical model admits as initial conditions both ${\dot R}>0$ 
and ${\dot R}<0$. By assuming from the start that $p_R>0$ they have
mutilated the original classical model,  unduly restricting  the allowed
set of initial conditions. Simply put, since the set of Hamilton's equations
generated by their   Hamiltonian (8) fails
to be  equivalent to the full set of classical equations of
motion, the conclusions based on its quantum version (9) are meaningless.

Contrary to what Gotay and Demaret impute us in their Comment, never do we
employ a ``modified quantum Hamiltonian'' or make ``ad hoc modifications to the
quantum dynamics.'' The Hamiltonian operator considered 
in [4] is {\it naturally} suggested by the
form of the {\it correct} reduced Hamiltonian (6), and it is as naturally 
a positive operator as the classical Hamiltonian (6) is a positive function. On
the other hand, the operator (9) {\it is simply not the quantum Hamiltonian}
because it is the quantum counterpart of the {\it incorrect} classical
Hamiltonian.
We maintain, therefore, that our paper [4] is correct and unequivocally
disproves Conjecture (F).

\vskip 1.25cm      

\noindent {\bf III. DUST-FILLED FRW MODELS}

\vskip .75cm

The model investigated in [2,3] is a FRW universe filled with dust. The extended
phase space is spanned by the canonical coordinates $(R,p_R)$ and 
$(\varphi ,p_{\varphi})$, where $\varphi$ is the only nonvanishing velocity 
potential for dust, with $p_{\varphi}>0$. The super-Hamiltonian constraint is

$$p_{\varphi} - \frac{p_R^2}{24R} - 6kR = 0 \,\,\, . \eqno(10)$$
\\
\noindent For $k=0$ or $k=-1$, if $p_R > -12kR$ the model expands from an
initial singularity, while if $p_R < 12kR$ it collapses to a final singularity.
In [3] the model is shown to be self-adjoint but nonsingular in the slow-time
gauge $t=p_R$, thus disproving Conjecture (S). Gotay and Demaret [1] object 
that the choice of time  $t=p_R$ is not permissible quantum mechanically.

It is true that Conjectures (F) and (S) were put forward only for ``dynamically
admissible'' choices of time. In [5] a gauge is said to be {\it dynamically
admissible} if the variable chosen as time is {\it a priori} bounded neither
above nor below. For the sake of simplicity, let us take the case $k=0$. As in
the previous model containing a scalar field, the reduced phase space of the
dust-filled model is disconnected into two components, corresponding to whether 
$p_R > 0$ or $p_R < 0$. Then, fixing one of these components, say 
$p_R < 0$, Gotay and Demaret state that ``$p_R $ is {\it a priori} bounded
above by zero.''

This argument suffers from the same type of deficiency as the one they used
regarding the scalar field model. Choices of time such as $t=R$ or
$t=p_R^2$ are true examples of dynamically inadmissible gauges, because in
these cases $t$ is {\it a priori} bounded below by zero, without the need of
unjustifiable additional restrictions on the classical model.
The canonical variable $p_R$ is not {\it a
priori} bounded above or below. It becomes bounded above or below only after
selecting one of the two components of the reduced phase space, and thus again
unduly restricting the set of allowed initial conditions, which amounts to
a mutilation of the original classical model.  Therefore, the
choice of time $t=p_R$ in [3] is dynamically admissible, and the same can be
said of the gauge $t=p_{\mu}$, with $\mu = \ln R$, employed in [2]. Thus, we
hold that papers [2] and [3] are correct and disprove Conjecture (S).

\vskip 1.25cm

\noindent {\large{\bf IV. CONCLUSION}}

\vskip .75cm                                                               

We have discussed quantum singularities on the basis of a singularity criterion
that involves the expectation value of an operator $\hat f$ whose classical
counterpart $f$ vanishes at the classical singularity [5], and certain issues
raised in [1] pertain more properly to the question whether this is a
reasonable criterion. The fact that in the dust model
one may have ${\hat R}(0) = 0$
without apparently anything catastrophic or pathologic happening to the quantum
system is a symptom that this criterion is not physically adequate.
It is also doubtful whether it is general enough to be applicable in all
gauges. It seems, therefore, 
that the formulation of
a general and physically reasonable singularity criterion remains an open
problem in quantum cosmology.

\newpage

\centerline{\bf REFERENCES}
\begin{description}

\item{[1]} M. J. Gotay and J. Demaret, {\it A Comment on Singularities in 
           Quantum Cosmology}, gr-qc/9605025.                                 
         
\item{[2]} N. A. Lemos, Phys. Rev. {\bf D41}, 1358 (1990).

\item{[3]} N. A. Lemos, Class. Quantum Grav. {\bf 8}, 1303 (1991).            
           
\item{[4]} N. A. Lemos, Phys. Rev. {\bf D53}, 4275 (1996).

\item{[5]} M. J. Gotay and J. Demaret, Phys. Rev. {\bf D28}, 2402 (1983).    

\item{[6]} W. F. Blyth and C. J. Isham, Phys. Rev. {\bf D11}, 768 (1975).

\end{description} 
\end{document}